\documentclass[aps,twocolumn,nofootinbib]{revtex4}
\usepackage{amsmath,epsfig}

\newcommand{\ie}{{\it i.e.}}
\newcommand{\eg}{{\it e.g.}}
\newcommand{\madgraph}{{\sc MadGraph}}
\newcommand{\feynrules}{{\sc Feyn\-Rules}}
\newcommand{\met} {\ensuremath{E\!\!\!\!/_T}}
\def\wt{\tilde}

\begin{document}

\leftline{}
\rightline{CERN-PH-TH/2013-229, DCPT/13/154, IPPP/13/77}

\title{
 Multilepton signals of gauge mediated supersymmetry breaking at the LHC
}

\author{
  Jorgen D'Hondt$^{a,b}$,
  Karen De Causmaecker$^{a,b,c}$,
  Benjamin Fuks$^{c,d}$,\\
  Alberto Mariotti$^e$,
  Kentarou Mawatari$^{a,b}$,
  Christoffer Petersson$^{b,f,g}$
  and Diego Redigolo$^{b,f}$
  }

\affiliation{{\phantom.}\\
$^{(a)}$ \mbox{Theoretische Natuurkunde and IIHE/ELEM, Vrije Universiteit Brussel,
  Pleinlaan 2, B-1050 Brussels, Belgium}\\
$^{(b)}$ \mbox{International Solvay Institutes, Brussels, Belgium}\\
$^{(c)}$ \mbox{Theory Division, Physics Department, CERN, CH-1211 Geneva 23, Switzerland}\\
$^{(d)}$ \mbox{Institut Pluridisciplinaire Hubert Curien/D\'epartement Recherches Subatomiques}, 
    Universit\'e de Strasbourg/CNRS-IN2P3, 23 Rue du Loess, F-67037 Strasbourg, France\\
$^{(e)}$ \mbox{Institute for Particle Physics Phenomenology, Durham University, Durham DH1 3LE, United Kingdom}\\
$^{(f)}$ \mbox{Physique Th\'eorique et Math\'ematique, Universit\'e Libre de Bruxelles,
  C.P. 231, B-1050 Brussels, Belgium}\\
 $^{(g)}$ \mbox{Department of Fundamental Physics, Chalmers University of Technology, 412 96 G\"oteborg, Sweden}
    }

\begin{abstract}
 We investigate multilepton LHC signals arising from electroweak
 processes involving sleptons. We consider the framework of general
 gauge mediated supersymmetry breaking, focusing on models where the
 low mass region of the superpartner spectrum consists of the three generations of
 charged sleptons and the nearly massless gravitino.
 We demonstrate how such models can provide an explanation for the anomalous
 four lepton events recently observed by the CMS collaboration, while
 satisfying other existing experimental constraints. The best fit to the 
 CMS data is obtained for a selectron/smuon mass of around 145~GeV and
 a stau mass of around 90~GeV. These models also give rise to final
 states with more than four leptons, offering alternative channels in
 which they can be probed and we estimate the corresponding production
 rates at the LHC.
\end{abstract}


\maketitle

\section{Introduction}

Despite the tremendous success of the Standard Model (SM) of particle
physics, the SM leaves many questions unanswered and it hints toward the
existence of new physics around the TeV scale. The arguably strongest
hint, which is reinforced by the recent observation of a Higgs boson at
the Large Hadron Collider (LHC)~\cite{Aad:2012gk,Chatrchyan:2012gu},
arises from the quadratic sensitivity of the Higgs mass parameter to
physics beyond the SM. This so-called hierarchy problem is addressed by
weak scale supersymmetry~\cite{Nilles:1983ge,Haber:1984rc}. Relating
fermionic and bosonic degrees of freedom, supersymmetry (SUSY) not only
stabilizes the weak scale but can also provide an explanation for
dark matter in the Universe and give rise to gauge coupling unification
at high energies. Consequently, searches for the superpartners of the SM
particles play key roles in the experimental program at the LHC.

The ATLAS and CMS collaborations have so far focused mainly on analyzing
signatures arising from the strong production of squarks and
gluinos. However, the negative search results have increased the interest
in analyses of the production of the electroweak
superpartners, whose cross sections are much smaller than colored ones.
Consequently, both collaborations have recently, for the first time,
been able to put
bounds on these particles that are stronger than those extracted from LEP data,
as shown, \eg, in Refs.~\cite{ATLAS-CONF-2013-049,CMS-PAS-SUS-13-006}.

In this letter we consider the framework of gauge-mediated SUSY breaking (see
Ref.~\cite{Giudice:1998bp} for a review and original references) in its
general formulation (GGM)~\cite{Meade:2008wd}, where it is possible to construct
models in which all the colored superpartners are heavy but some
(or all) of the electroweak superpartners are light. One benefit of this
kind of spectrum is that a 125 GeV Higgs boson can be easily
accommodated by means of multi-TeV top squarks.

We focus on models in which the three generations of right-handed
sleptons, together with the nearly massless gravitino, are in the low mass region
of the superpartner spectrum. Such models can be probed at the LHC by
analyzing events originating from the pair-production of sleptons that
decay promptly into lepton-rich final states with missing transverse
energy $\met$ carried by gravitinos.

We show that some of these GGM models can provide an explanation for a possible
anomalous production of events with four leptons recently observed by
the CMS collaboration~\cite{SUS-13-002}. We also discuss the compatibility
with the constraints extracted from the
dilepton$+\met$ searches at both LEP and LHC experiments,
as well as from other LHC multilepton searches. We finally propose,
for the model that fits the data best, additional signatures that could
be searched for using both data from the previous LHC runs and
future data from the run at a center-of-mass energy of $\sqrt{s}=13$~TeV.

\section{Theoretical framework and benchmark scenarios}

We consider a class of GGM models where the selectron and smuon
(generically referred to as sleptons in the following), as well as the
stau, lie in the low-mass range of the superparticle spectrum.
As for any scenario with gauge-mediated SUSY breaking~\cite{Giudice:1998bp},
the lightest supersymmetric particle (LSP) is the gravitino,
whose typical mass is ${\cal O}$(eV) for SUSY-breaking scales of 
${\cal O}$(100 TeV).

Adopting a bottom-up approach for new physics, we investigate the
phenomenology of a simplified model in which we extend the SM field
content by adding a nearly massless gravitino $\tilde{G}$, a pair of
mass-degenerate right-handed sleptons 
$\tilde\ell_R=\tilde e_R,\tilde\mu_R$ and a (for simplicity, non-mixed)
stau $\tilde \tau_R$. In addition, we also include the lightest 
neutralino state, considered to be bino-like and heavier than both the
sleptons and the stau. All the
remaining superpartners are assumed heavy and
effectively decoupled. 
Similar scenarios were considered in Refs.~\cite{Ambrosanio:1997bq,Ruderman:2010kj}.

\begin{figure}[t]
\centering
 \includegraphics[width=.4\textwidth]{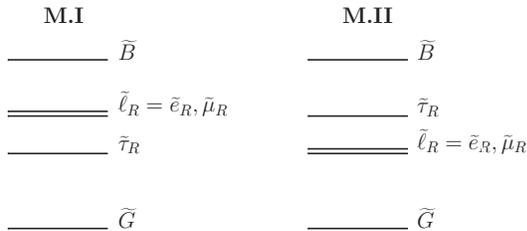}
 \caption{\label{fig:models} Mass spectra for our simplified model of
 type {\bf M.I} (left) and {\bf M.II} (right). In {\bf M.I} scenarios, 
 the stau $\tilde\tau_R$ is the NLSP and the right-handed
 selectron/smuon $\tilde\ell_R$ are co-NNLSP. In models of class
 {\bf M.II}, the situation is reversed.} 
\end{figure}

In this simplified model, two possible hierarchies can be realized in
the slepton/stau sector. As presented in Fig.~\ref{fig:models}, we
consider both of these and denote by {\bf M.I} scenarios where
the stau is the next-to-lightest superparticle (NLSP) and sleptons
the next-to-next-to-lightest superpartners (NNLSP), and by {\bf M.II}
scenarios with an inverted hierarchy, with the sleptons being co-NLSP
and the stau the NNLSP. While models of type {\bf M.I} are typical
in GGM (even in minimal gauge mediation), models of type {\bf M.II}
can be realized when the soft masses for both Higgs fields at
the UV scale are allowed to receive extra, non-gauge mediated,
contributions~\cite{Evans:2006sj,Grajek:2013ola}. 

\begin{figure}
\centering
 \includegraphics[width=.33\textwidth]{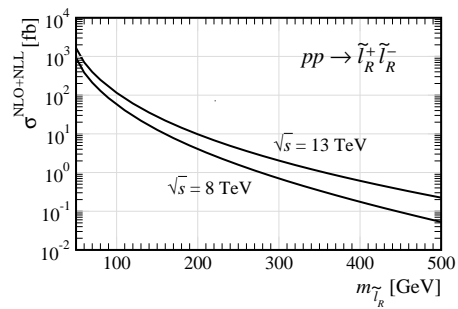}
 \caption{\label{fig:xsec} Right-handed slepton/stau pair-production cross
 section at the LHC, for a single flavor, as a function of the slepton mass.}
\end{figure}

Slepton pairs are produced via the electroweak Drell-Yan process. Due to
the steeply falling cross section with increasing slepton
masses~\cite{Beenakker:1999xh}, we consider slepton and stau masses only
up to 300~GeV, a range above which it is unlikely that the LHC at 
$\sqrt{s}=8$~TeV is sensitive. This is illustrated in
Fig.~\ref{fig:xsec} where we present the production cross section
of a right-handed slepton/stau pair at the LHC, for
$\sqrt{s}=8$~TeV and 13~TeV, as computed by
{\sc Resummino}~\cite{Bozzi:2006fw,Bozzi:2007qr,Bozzi:2007tea,Fuks:2013vua}. 

\begin{figure}
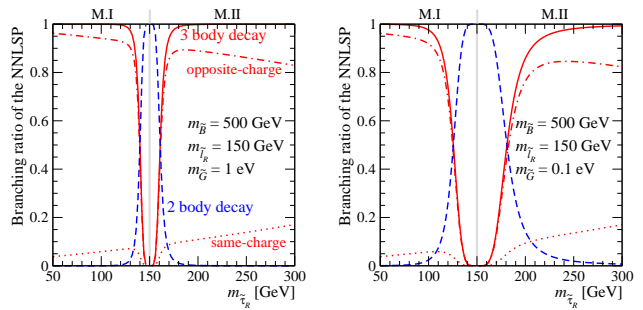

\centering
\vspace{.24cm}
 \includegraphics[width=.22\textwidth]{br1}
 \hspace{.3cm}\includegraphics[width=.22\textwidth]{br2}
 \caption{\label{fig:BR} Branching ratio of the NNLSP as a function of
 the stau mass for two different choices of the gravitino mass,
 $m_{\tilde G}=1$~eV (left) and $0.1$~eV (right), where the bino and
 slepton masses are fixed at 500~GeV and 150~GeV, respectively. The
 dashed red line corresponds to the two-body decay to the gravitino, while
 the solid blue line indicates the total three-body decay branching
 ratio. The red dashed-dotted and dotted lines represent the
 opposite-charge and same-charge three-body decays, respectively, as
 explained in the text.}
\end{figure}

For both types of scenarios, the NLSP universally decays into a
gravitino and the corresponding SM partner,
\begin{align}
 \tilde\tau_R\to\tau\tilde G\ \ ({\rm\bf M.I});\qquad
 \tilde\ell_R\to\ell\tilde G\ \ ({\rm\bf M.II})\ ,
\end{align}
with a decay length depending on the gravitino mass~\cite{Giudice:1998bp}.
We require this decay to be prompt so that an upper bound
on the gravitino mass is imposed at around $10$~eV.

Concerning the NNLSP, the analogous two-body decay
competes with possible three-body decay modes via an off-shell bino,
\begin{align}
 \wt{\ell}_R\to\ell\tau\wt{\tau}_R\ \ ({\rm\bf M.I});\qquad
 \wt{\tau}_R\to\tau\ell\wt{\ell}_R\ \ ({\rm\bf M.II})\ .
\label{3bd}
\end{align}
Fig.~\ref{fig:BR} presents,
for two different choices of the
gravitino mass $m_{\tilde G}=1$~eV (left) and $0.1$~eV (right),
the NNLSP two-body and three-body branching
ratios when fixing the bino mass to 500~GeV, the slepton mass
$m_{\tilde\ell_R}$ to 150~GeV
and when varying the stau mass $m_{\tilde\tau_R}$.  When
$m_{\tilde\tau_R}<m_{\tilde\ell_R}$ ({\bf M.I}), we display the decay
modes of the slepton, whereas when
$m_{\tilde\tau_R}>m_{\tilde\ell_R}$ ({\bf M.II}) the ones of the
stau. The three-body decay is found
dominant except in the region where the NNLSP and NLSP are close in mass
($m_{\tilde\tau_R}\approx m_{\tilde\ell_R}$). This result is robust
under variations of the bino mass.
Our models exhibit a suppression of the two-body decay mode of the
NNLSP into the gravitino LSP by the SUSY-breaking scale. As will be shown below,
this is a key feature to get agreement with data. Nevertheless, it is possible
that such a suppression could be achieved in other scenarios. For instance, one
could try to replace the gravitino with a singlino. Again, the question would be
how to accommodate a prompt two-body decay of the NLSP and, simultaneously, a
dominant three-body decay of the NNLSP. One possibility (at least for stau NLSP
models) might be to use singlino mass mixing with higgsinos and the hierarchies
among the lepton Yukawa couplings. However, this goes beyond the scope of
the present paper in which we focus on GGM models for the Minimal Supersymmetric
Standard Model only.

We briefly notice that the three-body decay distinguishes between the
different charge channels, \ie, the NLSP can have either the opposite
charge of the NNLSP ($\wt{\ell}_R^-\to\ell^-\tau^-\wt{\tau}_R^+$) or the
same 
($\wt{\ell}_R^-\to\ell^-\tau^+\wt{\tau}_R^-$)~\cite{Ambrosanio:1997bq},
denoted by dashed-dotted and dotted lines in Fig.~\ref{fig:BR}. 
Generically, the more the bino is off-shell, the more the opposite
charge channel dominates. 
Since the dominance of one channel with respect to the other is very
much dependent on whether the sleptons are right- or left-handed, on
the amount of stau mixing and on the nature of the neutralino, a detailed analysis of these effects might give us
a way of probing non-trivial properties of the spectrum. However, the
current LHC statistics is too low to allow for this
analysis that we leave for further investigation.

\section{Multilepton signals in gauge mediated supersymmetry breaking}

Recently the CMS collaboration reported a slight excess in events with
three electrons or muons (out of which one opposite-sign same flavor lepton pair can be formed)
and one hadronically decaying tau, in the category with a $Z$-veto, low hadronic activity and
no jet issued from the fragmentation of a $b$-quark~\cite{SUS-13-002}.
With 19.5~fb$^{-1}$ of collisions at $\sqrt{s}=8$~TeV,
the number of observed (expected) events in this category is 15 (7.5$\pm$2), 4 (2.1$\pm$0.5) and 3 (0.6$\pm$0.24) for the three 
regions $\met<50$~GeV,
$\met \in [50,100]$~GeV and
$\met>100$~GeV, respectively.

\begin{figure}[t]
\centering
 \includegraphics[width=.238\textwidth]{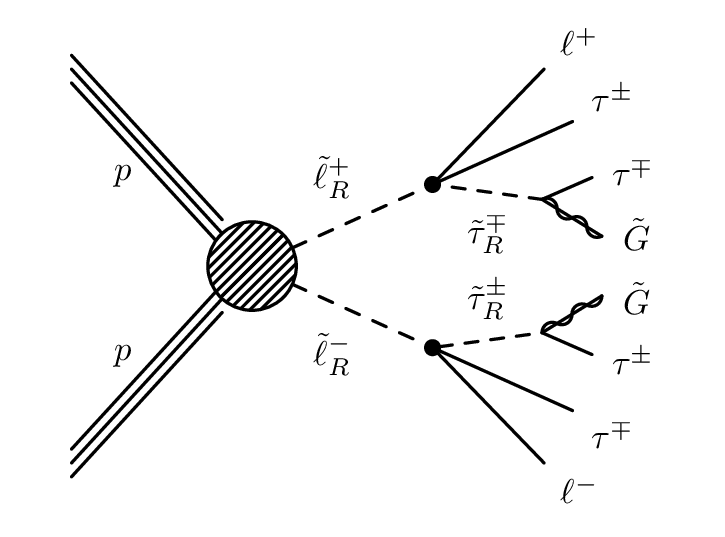}
 \includegraphics[width=.238\textwidth]{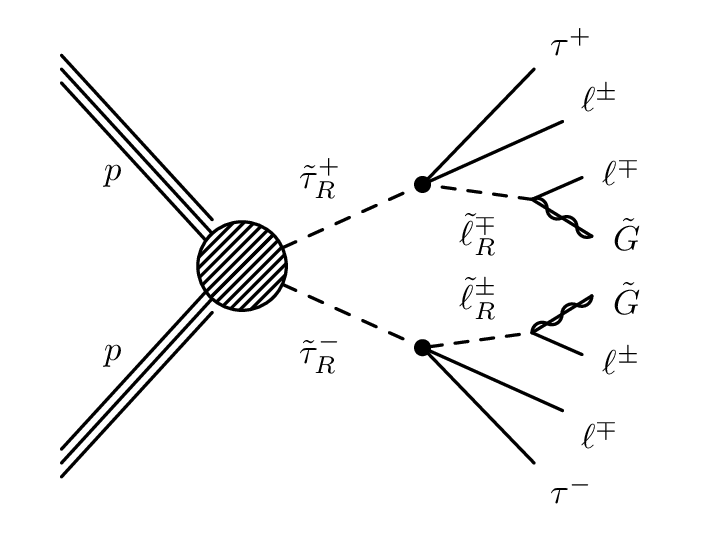}
 \caption{\label{fig:feynman} Diagrams leading to multilepton production
 in association with missing energy in scenarios of type {\bf M.I} (left) and
 {\bf M.II} (right).} 
\end{figure}

Motivated by this result, we investigate
the contributions arising from slepton and stau
pair production for models of class {\bf M.I} and {\bf M.II}. To display our
results, we fix the bino and gravitino masses to 500~GeV and 1~eV,
respectively, and scan the slepton and stau masses from 50~GeV to
300~GeV. Within our choice of parameters, the NNLSP dominantly decays
via its three-body mode in most
of the ($m_{\wt{\ell}_R},m_{\wt{\tau}_R}$) mass plane. This allows for a
possible enhancement of the production rates of final states
comprised of $4\tau+2\ell+\met$ and $2\tau+4\ell+\met$ for {\bf M.I} and
{\bf M.II} scenarios,
respectively, as depicted in Fig.~\ref{fig:feynman}. The actual
final state lepton multiplicity however depends on the number of
leptonically decaying taus.

For our SUSY signal simulation, we use the goldstino
model~\cite{Argurio:2011gu,Mawatari:2011jy}
implemented in the
\feynrules~package~\cite{Christensen:2008py,Duhr:2011se} and export it
to a UFO library~\cite{Degrande:2011ua} which has been linked to
\madgraph~5~\cite{Alwall:2011uj}. The generated parton-level events
have then been processed by
{\sc Pythia}~\cite{Sjostrand:2006za} for parton showering and hadronization,
{\sc Tauola}~\cite{Jadach:1993hs} for tau decays and by
{\sc Delphes}~\cite{Ovyn:2009tx} for detector simulation using the recent
CMS detector description of Ref.~\cite{Agram:2013koa}. We have
analyzed 19.5~fb$^{-1}$ of events describing
NNLSP pair production at the LHC, running at $\sqrt{s}=8$~TeV,
with {\sc MadAnalysis~5}~\cite{Conte:2012fm}. Generated events have
been reweighted using signal cross sections
predicted by {\sc Resummino} at the
next-to-leading order and next-to-leading logarithmic
accuracy, as shown in Fig.~\ref{fig:xsec}. This results
in typical $K$-factors
of about $1.2$ for the scanned mass range.

For event selection, we follow the
CMS multilepton analysis of Ref.~\cite{SUS-13-002} and base
our results on an investigation of the properties of
isolated electron and muon candidates whose transverse-momentum
$p_T$ is greater than 10~GeV and pseudorapidity $|\eta|$ is smaller
than 2.4.
We enforce lepton isolation by imposing the amount of transverse
activity in a cone of radius $R \!=\! \sqrt{\Delta\varphi^2 +
\Delta\eta^2} \!=\! 0.3$ centered on the lepton, $\varphi$ being the azimuthal angle with
respect to the beam direction, to be less than 15\% of the lepton $p_T$.
Additionally, we impose the
leading lepton (electron or muon) transverse momentum to satisfy $p_T>20$~GeV
and include efficiencies of 95\%, 93\%
and 90\% to simulate the effects of
the double-electron, electron-muon and double-muon
triggers relevant for the considered final state topologies.
Finally, events featuring a pair of opposite-sign same flavor (OSSF)
leptons whose invariant-mass is smaller than
12~GeV are rejected.
While leptonically-decaying taus are
accounted for as the electrons or muons in which they decay into,
hadronically-decaying taus $\tau_h$ are reconstructed as such
and we demand their visible $p_T$ to be greater than 20~GeV
and their pseudorapidity to fulfill $|\eta|<2.3$.

\begin{figure*}
\centering
 \includegraphics[width=.245\textwidth]{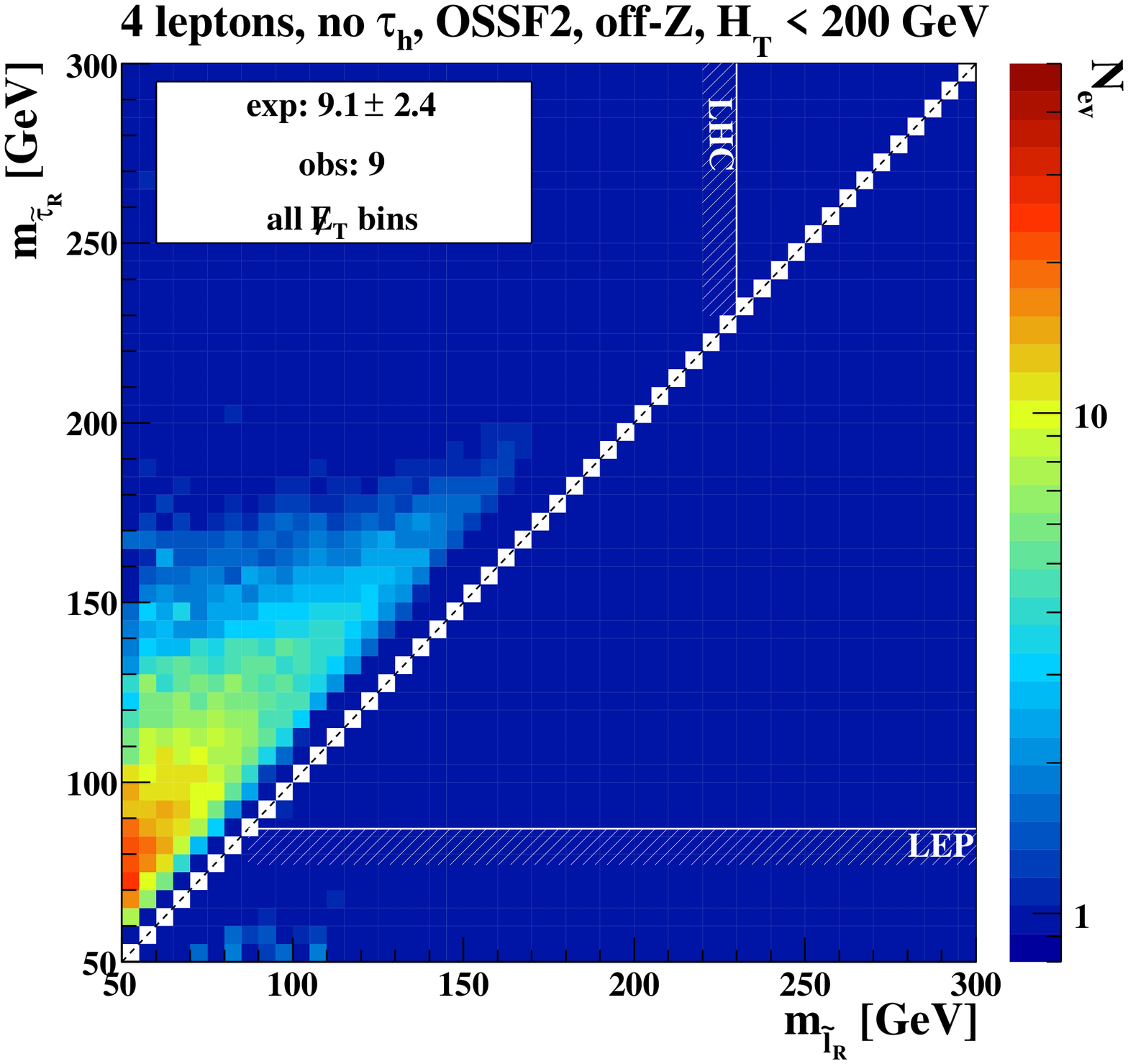}
 \includegraphics[width=.245\textwidth]{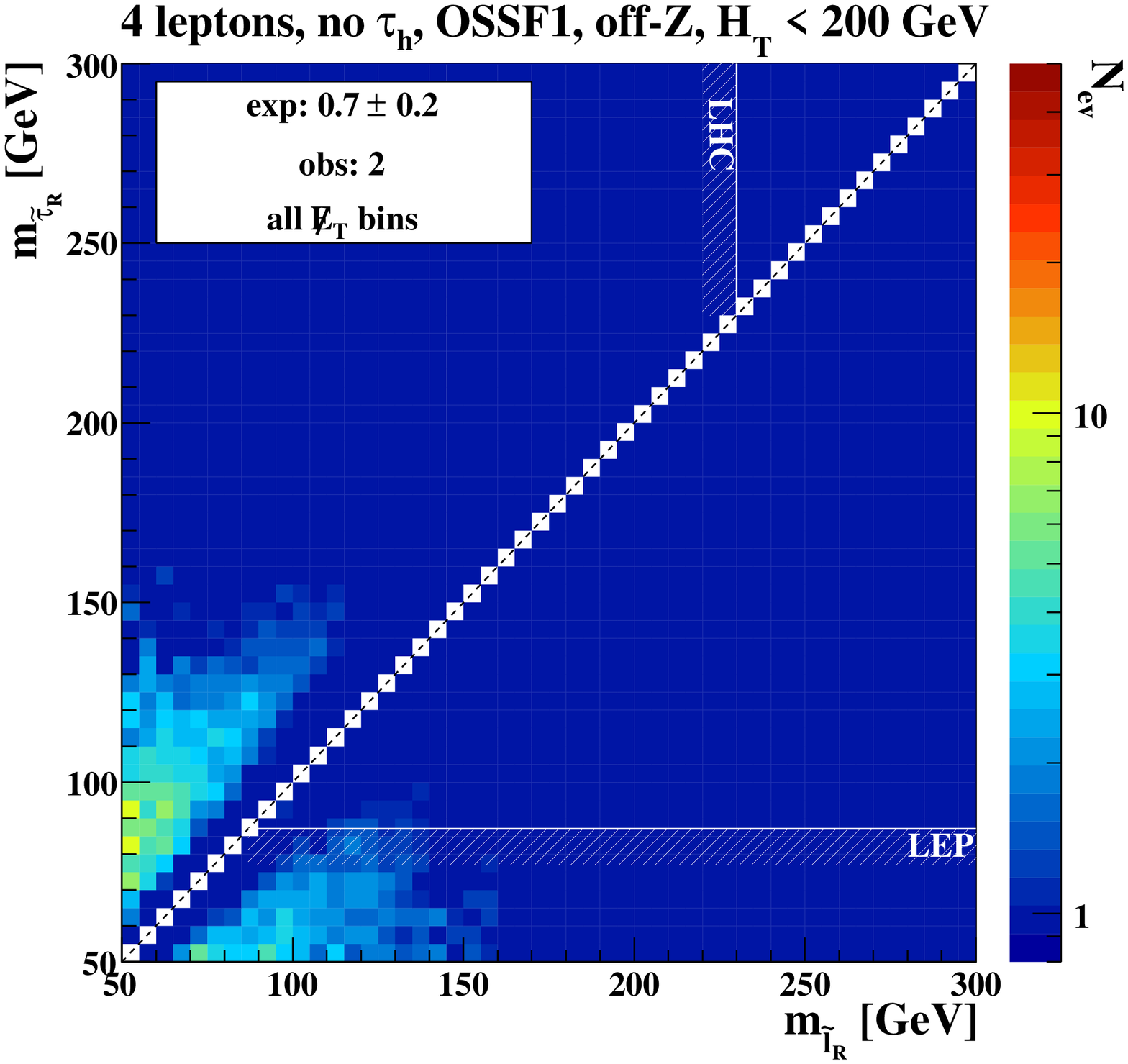}
 \includegraphics[width=.245\textwidth]{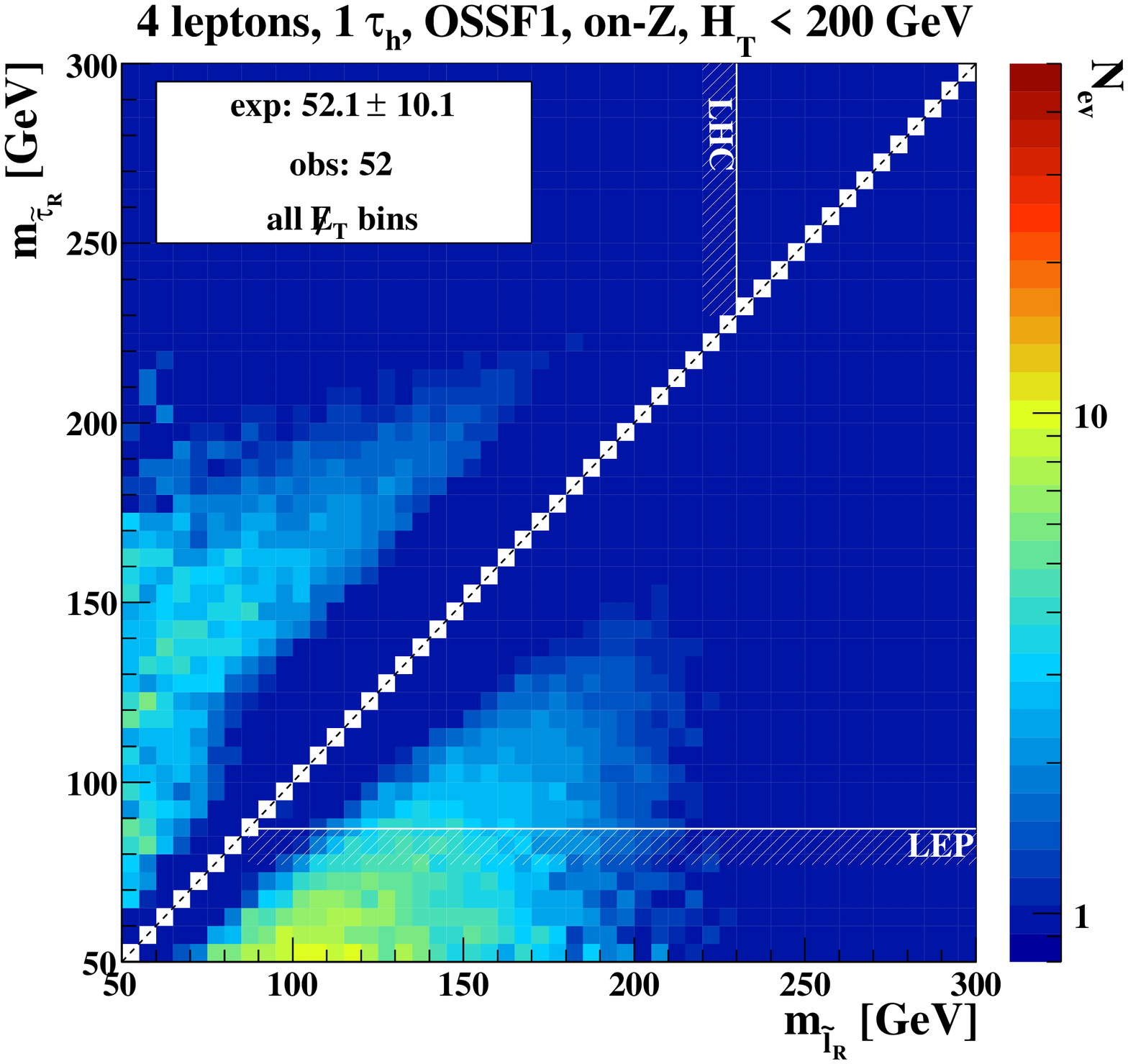}
 \includegraphics[width=.245\textwidth]{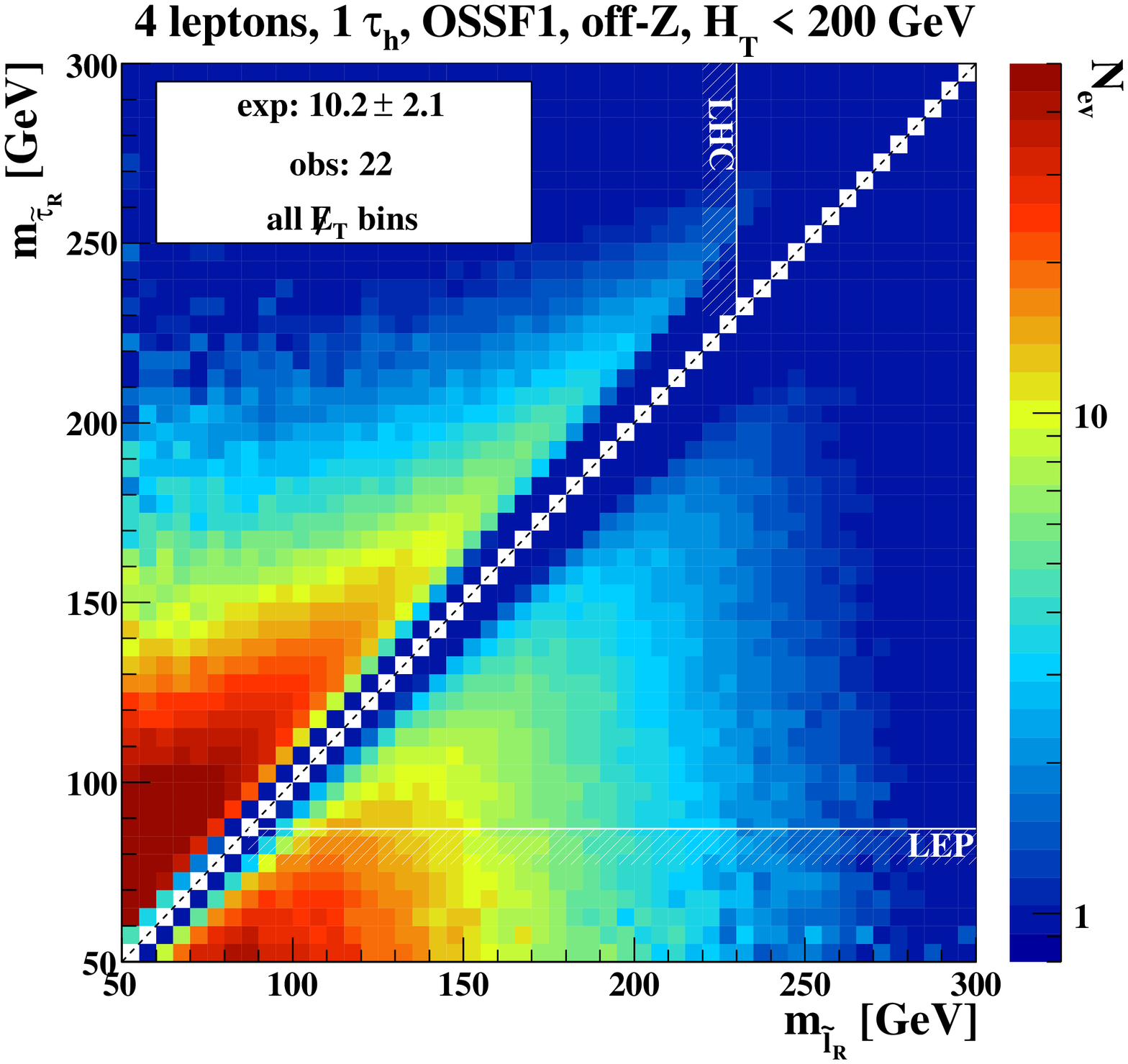}
 \caption{\label{fig:CMS1} Number of the signal event in the 
 ($m_{\wt{\ell}_R},m_{\wt{\tau}_R}$) plane for some representative
 categories, after summing $\met$ bins. The numbers of
 expected and observed events
 are also presented~\cite{SUS-13-002}, as well as LEP and LHC bounds from direct slepton
 searches.} 
\end{figure*}

The CMS analysis classifies events as having $H_T$ greater or less than
200~GeV as well as counting the number of $b$-tagged jets in the final states
for which we employ the $b$-tagging algorithm described in Ref.~\cite{Agram:2013koa}.
The $H_T$ variable is defined as the
scalar sum of the transverse energy of
all isolated reconstructed jets (not including hadronic tau contributions)
with $p_T>30$~GeV and $|\eta|<2.5$, for which
we use an anti-$k_T$ algorithm
whose radius parameter is fixed to $R=0.5$~\cite{Cacciari:2008gp},
as implemented in the {\sc FastJet} package~\cite{Cacciari:2011ma}, and we consider
a jet as isolated only if no electron, muon or tau lies within a cone of
radius $R=0.3$ centered on the jet.
Concerning signal events, the hadronic activity mainly
arises from initial state radiation so that $H_T$
is always found smaller than 200~GeV and the number of $b$-jets
is rarely above zero. This feature is actually welcome since the CMS
experiment does not see any excess in the regions where
$H_{T}>200\text{ GeV}$ or $N_{\text{b-jets}}\geq 1$.

After applying the
above requirements, events with at least three leptons are selected, where at
most one of them is a hadronic tau.
Further categories are made by classifying each event in terms of the
maximum number of opposite-sign same flavor (OSSF) lepton pairs.
Final state signatures predicted by both {\bf M.I} and {\bf M.II} models
contain at least one
OSSF lepton pair in most of the parameter space, which is again a
welcome feature since the bins with zero OSSF lepton pairs do not
exhibit any excess.
The `on-$Z$' region is populated if at least one
OSSF lepton pair has an invariant mass in the $Z$-window
$|m_{\ell^+\ell^-}-m_Z|<15$~GeV while in the `off-$Z$' region,
each OSSF dilepton invariant mass lies
outside the $Z$-window.

After summing the $\met$ bins, we have six categories for both the four lepton and the
three lepton cases. We focus our discussion mainly on the
four lepton channels since in the three leptons ones, the expected background
is so large that the contributions from our signal region, characterized
by a small yield, are always in agreement with the expectation within
the statistical precision.

For illustrative purposes, in Fig.~\ref{fig:CMS1} we show four
categories out of the possible six  for the four lepton case, displaying the
number of signal events in the ($m_{\wt{\ell}_R},m_{\wt{\tau}_R}$) mass
plane. We also quote the numbers of expected and observed events from
Table 2 in the CMS note~\cite{SUS-13-002}. 
The lower half plane, with $m_{\wt{\tau}_R}<m_{\wt{\ell}_R}$,
corresponds to the {\bf M.I} models, while the upper half plane, with
$m_{\wt{\tau}_R}>m_{\wt{\ell}_R}$, corresponds to the {\bf M.II} models.

In the category with two `off-$Z$' OSSF lepton pairs,
corresponding to the first panel of Fig.~\ref{fig:CMS1},
the CMS analysis finds good agreement with the SM expectation. While models of class
{\bf M.I} do not give rise to any signal events in this category,
the {\bf M.II} models that are best compatible
with the data in this category are those with
$m_{\tilde{\tau}_{R}}\gtrsim150\text{ GeV}$, \ie\, those that give rise
to very few signal events.
The category with one `off-$Z$' pair of OSSF leptons and no hadronic tau is shown as the second panel
of Fig.~\ref{fig:CMS1}. For very low stau masses, {\bf M.I} scenarios can populate this bin
with events featuring at least two leptonically decaying taus. By comparing with the first panel of the figure,
we observe that out of the four leptons, {\bf M.I} models generally predict, in the absence of
hadronic taus, that one single OSSF lepton pair can be formed, whereas two OSSF lepton
pairs are rather expected in {\bf M.II} models.
In the third panel of Fig.~\ref{fig:CMS1}, we turn to the four lepton category including one hadronic tau
and where one single OSSF lepton pair can be formed and lies in the $Z$-window. All scanned
{\bf M.I} and {\bf M.II} scenarios predict number of events lying
comfortably within $1\sigma$ variation of the SM expectation.
The last panel of Fig.~\ref{fig:CMS1} shows the four lepton category including one hadronic tau
and one OSSF lepton pair whose invariant mass is not compatible with the $Z$-boson mass. This category
corresponds to the observed excess and both types of signal scenarios can provide good candidates
for explaining it.

\begin{figure*}[t]
\centering
 \includegraphics[width=.25\textwidth]{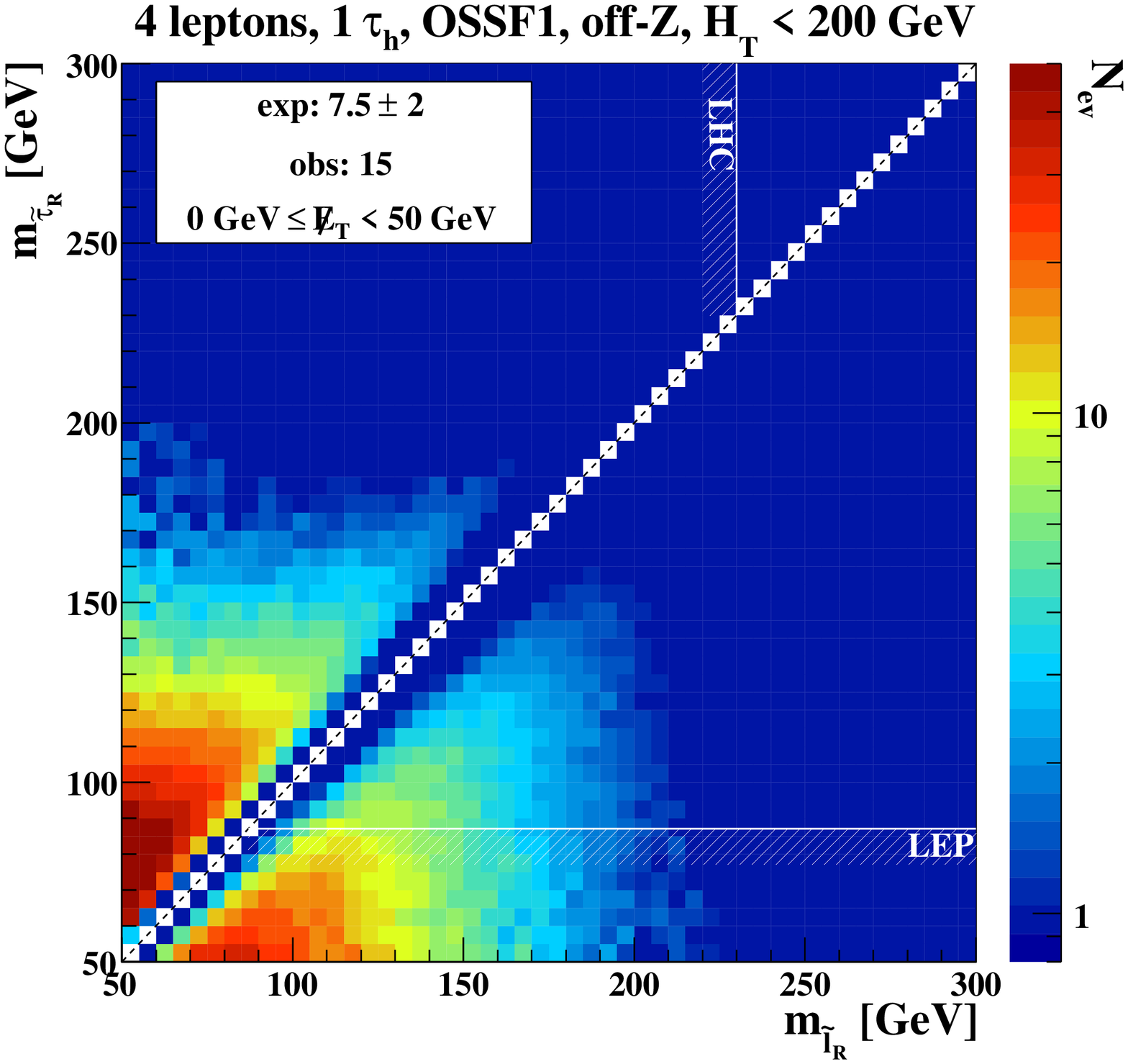}
 \includegraphics[width=.25\textwidth]{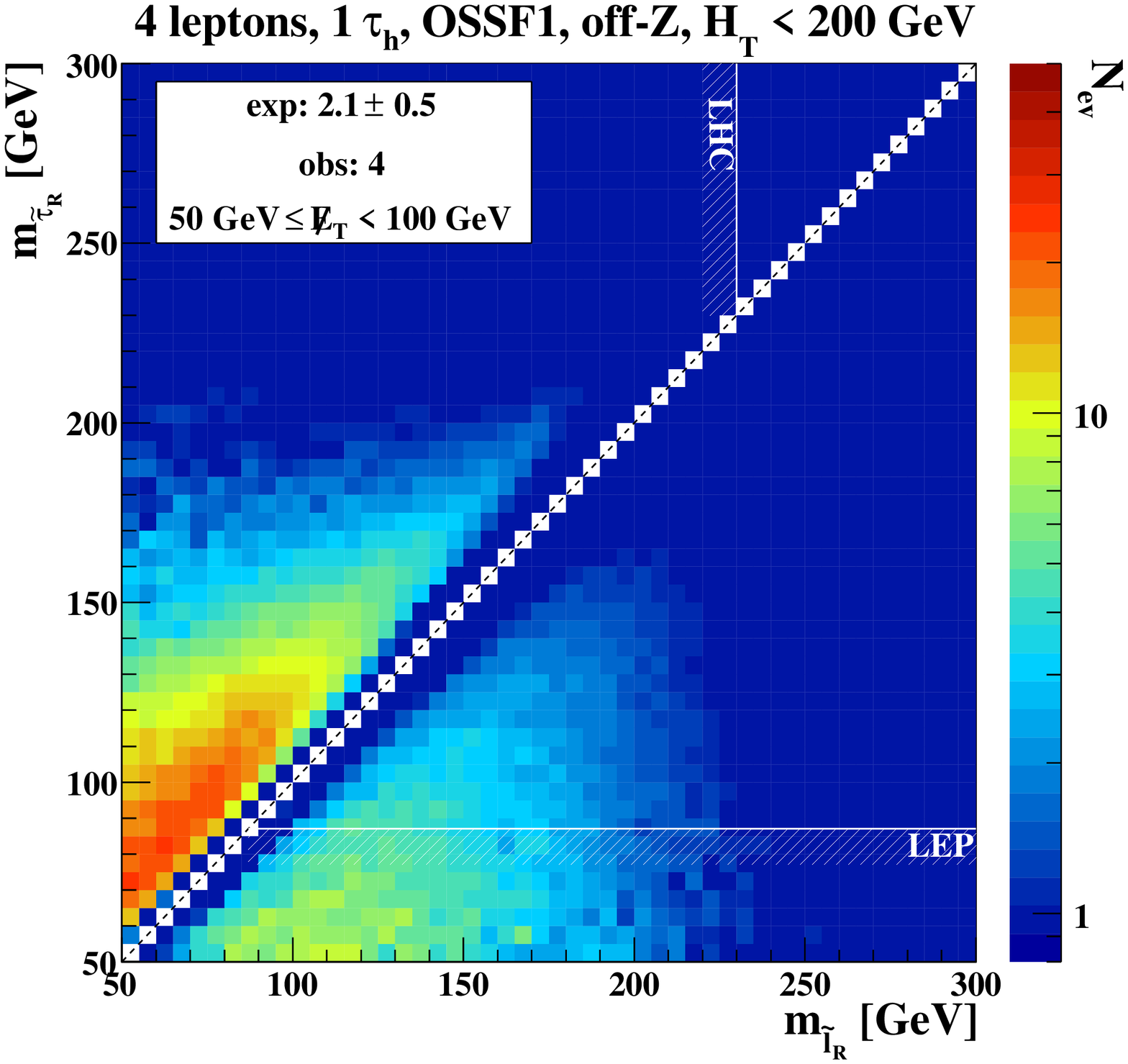}
 \includegraphics[width=.25\textwidth]{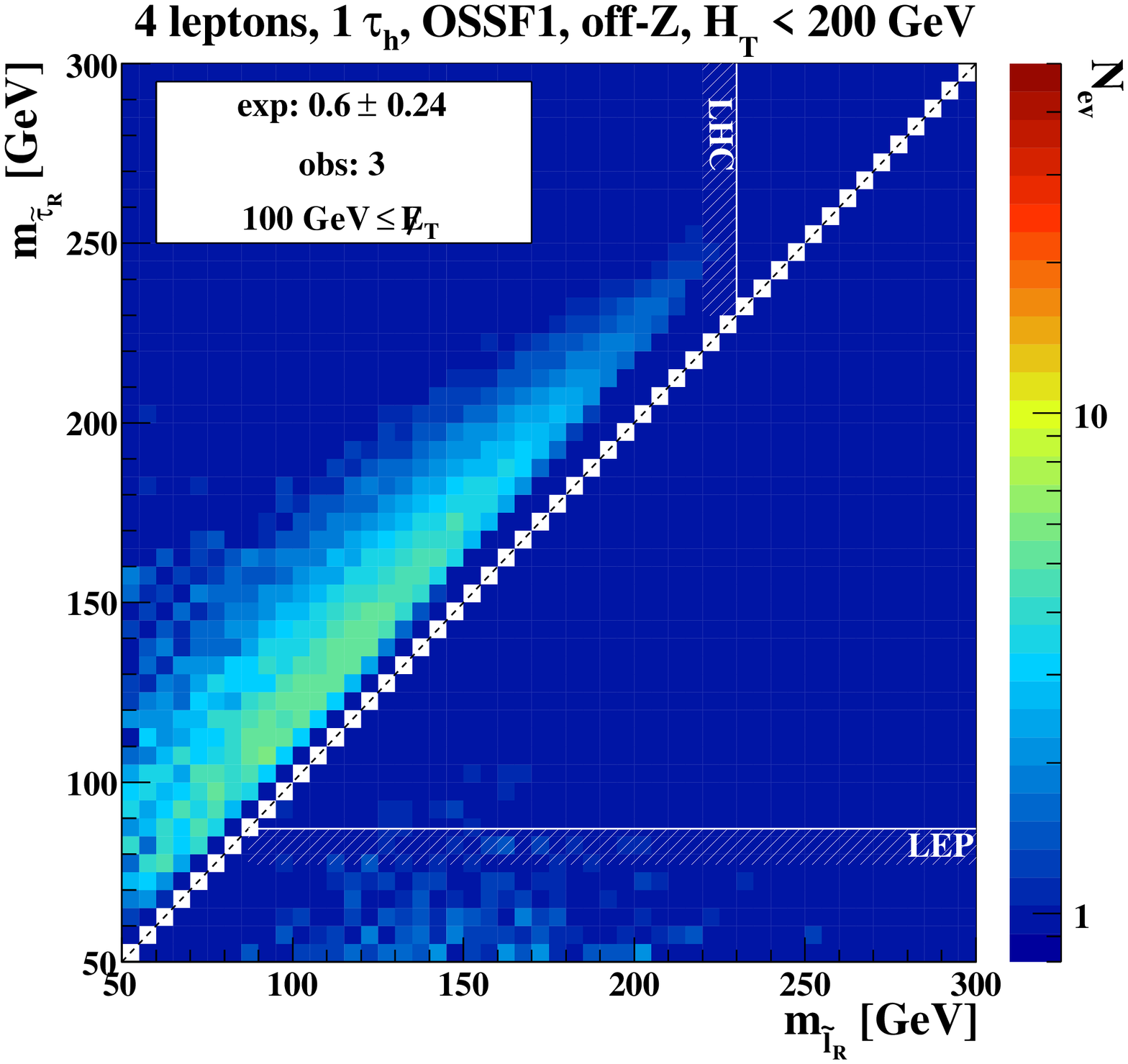}
 \caption{\label{fig:CMS2} The same as in Fig.~\ref{fig:CMS1}, but for a
 category where the excess events are observed and for different $\met$
 bins.} 
\end{figure*}

In Fig.~\ref{fig:CMS2}, we display the precise distribution of our signal in
the different $\met$ bins corresponding
to the last panel of Fig.~\ref{fig:CMS1}. Scenarios of class {\bf M.I}
do not populate the bin with $\met>100\text{ GeV}$,
unless in a narrow region where the stau is very light.
Performing a $\chi^2$ fit restricted to the three bins displayed in the figure for both class of
models, the best benchmark scenarios are given by
\begin{align*} 
& \text{{\bf M.I}}:\ m_{\tilde{\ell}_{R}}=140\text{ GeV},\
 m_{\tilde{\tau}_{R}}=50\text{ GeV},\ \ \chi^2_{\text{exc.}}=1.22\ ;\\
& \text{{\bf M.II}}: m_{\tilde{\ell}_{R}}=50\text{ GeV},\ \ m_{\tilde{\tau}_{R}}=140\text{ GeV}, \ \chi^2_{\text{exc.}}=2.28\ .
\end{align*}
Both models end up providing an explanation for the excess.
However, as detailed below, experimental constraints arising from direct NLSP
pair production exclude all {\bf M.II} candidates explaining the excess,
and have also non-trivial consequences on the best fit for {\bf M.I} models.

In {\bf M.II} models, where the right-handed sleptons are co-NLSP, current bounds on the
slepton mass apply, $m_{\wt\ell_{R}}>230$~GeV~\cite{ATLAS-CONF-2013-049,CMS-PAS-SUS-13-006}.
These bounds are extracted from slepton pair
production and subsequent decay into a lepton and a gravitino
(a nearly massless LSP).
As indicated in both Figs.~\ref{fig:CMS1} and~\ref{fig:CMS2}, this
excludes the entire region of the {\bf M.II} parameter space possibly relevant for explaining the CMS excess.
On the other hand, for {\bf M.I} scenarios in which the right-handed stau is the NLSP,
the most stringent constraints are those set by LEP experiments, $m_{\wt{\tau}_R}>87$~GeV~\cite{Abbiendi:2005gc},
as the corresponding LHC searches have a too low sensitivity~\cite{ATLAS-CONF-2013-028}.
Consequently, {\bf M.I} models still provide viable candidates for explaining the excess.

The point of the {\bf M.I} parameter space ending up to be the best fit of
the three bins with the excess becomes,
after accounting for LEP limits on the stau mass,
\begin{equation*}
m_{\tilde{\ell}_{R}}=145\text{ GeV},\ m_{\tilde{\tau}_{R}}=90\text{ GeV},\ \chi^2_{\text{exc.}}=2.42\ ,
\end{equation*}
where $m_{\tilde{\tau}_R}$ lies at the edge of the excluded region.
The significance of our best fit scenario is found reduced as signal contributions
to the low missing energy bin of Fig.~\ref{fig:CMS2} are smaller for larger stau masses.
This further motivates us to study in detail how and whether LHC direct searches could
improve LEP limits on the stau mass~\cite{ToAppear}.
As a crosscheck of our reasoning, we perform a global fit on the {\bf M.I} parameter space
including all four leptons categories. Not surprisingly,
the same best fit benchmark point
with $m_{\tilde{\tau}_{R}}=90\text{ GeV}$ and $m_{\tilde{\ell}_{R}}=145\text{ GeV}$
is obtained.

Focusing from now on on the best fit point, we briefly comment on other
signatures that it induces and which could be probed through
other multilepton searches at the LHC. Firstly, CMS searches for
$R$-parity violating (RPV) SUSY in leptonic final states
are not expected to be sensitive to such models as it requires
four electron or muons in the final states~\cite{CMS-PAS-SUS-13-010}. Such a
signature is suppressed in the framework of {\bf M.I} models (as already shown on the
first panel of Fig.~\ref{fig:CMS1})
as it requires at least two of the taus to decay leptonically.

Secondly, the ATLAS collaboration has recently performed a multilepton search
which features one signal region, dubbed `SR1noZ',
that might be relevant for models of class {\bf M.I}~\cite{ATLAS-CONF-2013-036}.
This analysis has been designed for RPV SUSY searches and requires exactly
three electrons or muons and at least one tau. An extended $Z$-veto is demanded
so that events with a lepton pair, triplet or quadruplet whose invariant mass
lies within a 20~GeV interval centered on the $Z$-boson mass are rejected.
The search strategy additionally requires either a selection on the
missing energy $\met>100\text{ GeV}$ or on the effective mass,
defined as the sum of the missing energy and of all the transverse momenta of the reconstructed
final state objects (leptons, hadronic taus, jets), $m_{\text{eff}}>400\text{ GeV}$.
On the one hand, our signal does not populate the $\met>100\text{ GeV}$ category as shown
on Fig.~\ref{fig:CMS2}. On the other hand, the tail of the effective mass distribution for our best
benchmark point has been found to only extend up to about 350~GeV,
which can be heuristically understood as most of the reconstructed final state objects come
from the decay of a slepton pair with an invariant mass of about $300\text{ GeV}$. 
This ATLAS search is therefore expected to be
insensitive to our benchmark.

\begin{table}
\begin{tabular}{c|c|cc}
\hline \hline
$N(\ell )$ & $N(\tau_h)$ & $~~N_{\text{events}}(8\text{ TeV})~~$ & $~~N_{\text{events}}(13\text{ TeV})~~$ \\ \hline \hline
4 & 2 & 22.5  & 223\\ \hline
5 & 0 & 0.074 & 0.79\\
5 & 1 & 1.7   & 14.7\\
5 & 2 & 7.4   & 76.1\\  \hline
6 & 0 & 0     & 0\\
6 & 1 & 0.075 & 0.66\\ 
6 & 2 & 1.0   & 7.89\\ \hline
$>6$ & 0 & 0.038 & 13.9\\ \hline
\end{tabular}
\caption{\label{TableI}Number of multilepton events $N_{\text{events}}$ predicted by the scenario
  that fits the CMS excess best ({\bf M.I} model, $m_{\tilde \ell}=145$ GeV, $m_{\tilde \tau} = 90$ GeV).
  The third column corresponds to 19.5~fb$^{-1}$ of LHC collisions at $\sqrt{s}=8$~TeV
  and the fourth column to 100~fb$^{-1}$ of LHC collisions at $\sqrt{s}=13$~TeV.
  Moreover, $N(\ell)$ denotes the total number of charged leptons and $N(\tau_h)$ how many
  of these are hadronically-decaying taus.}
\end{table}

Lastly, the ATLAS collaboration has recently performed an
investigation of ditau events~\cite{ATLAS-CONF-2013-028},
making use of a dedicated trigger on two
reconstructed hadronic taus. This analysis could be relevant
in our case since the signal is likely to populate bins with
two hadronic taus, as shown in Table~\ref{TableI}.
However, these taus are always accompanied by extra electrons
or muons issued from NNLSP three-body decays,
so that no hint in the ATLAS signal regions, which also
include a veto on additional leptons, is foreseen.

Let us finally discuss how some of
the existing searches can be optimized to improve their
sensitivity for signal scenarios of class {\bf M.I}.
As an illustrative example, we show in Table~\ref{TableI} that
our best fit point is considerably contributing to final
states with two hadronically decaying taus plus either two or three
electrons or muons. In particular, predicting a considerably large
number of events featuring three electrons or muons shows that the lepton abundance
in the final state can be considerably enhanced by the leptonically decaying taus, even
though the associated branching fraction is reduced.
For these reasons we point out that an optimized search strategy for {\bf M.I} models
should impose selections on the lepton multiplicity as inclusive as possible, as
already suggested in the context of optimizing Tevatron searches for gauge mediation
scenarios~\cite{Ruderman:2010kj}. Moreover, one peculiar feature of our benchmark scenario
is the presence of at least two hadronically decaying taus
which are hard enough to be reconstructed. We therefore suggest an effective search
dedicated to {\bf M.I} scenarios that
could be made by combining triggers on two hadronically
decaying taus with a binning on the number of extra leptons in the final state.

\section{Conclusions}
In this letter, we have investigated the phenomenology of models
involving light charged sleptons, realized
within the framework of general gauge mediated supersymmetry breaking.
Motivated by the recent CMS observation of an excess
in multilepton events, we have demonstrated that some of these
models can not only provide an explanation for the excess but also explain
why no hint of new physics has been found in other leptonic searches by
both the ATLAS and CMS collaborations.
We have shown that the model that
best fits the data, and which is
compatible with all current experimental constraints,
involves right-handed selectrons and
smuons of 145~GeV and a right-handed stau of 90~GeV. 
The presence of a light stau at the edge of the LEP limit in our benchmark motivates further investigation about the possible impact of LHC searches on the stau mass bound~\cite{ToAppear}.
Finally, we proposed
new investigations in multileptonic channels that could probe this type
of GGM models and further constrain them in the future.\\

\begin{acknowledgments}
The work of B.F. has been partially supported by the Theory-LHC France-initiative of the CNRS/IN2P3 and by
the French ANR 12 JS05 002 01 BATS@LHC.
J.D.H., K.D.C. and K.M. are supported in part by the Belgian Federal Science Policy Office through the Interuniversity Attraction Pole P7/37 and in part by the Strategic Research Program ``High Energy Physics" and the Research Council of the Vrije Universiteit Brussel.
K.D.C is also partially supported by a ``FWO-Vlaanderen'' aspirant fellowship and
acknowledges the hospitality of the CERN TH group and the support from the ERC grant 291377 ``LHCtheory: Theoretical predictions and analyses of LHC physics: advancing the precision frontier''.
The work of C.P. and D.R was supported in part by IISN-Belgium (conventions 4.4511.06, 4.4505.86 and 4.4514.08), by the ``Communaut\'e Fran\c{c}aise de Belgique" through the ARC program and by a ``Mandat d'Impulsion Scientifique" of the F.R.S.-FNRS. C.P. is supported by the Swedish Research Council (VR) under the contract 637-2013-475. A.M. acknowledges funding by the Durham International Junior Research Fellowship.
\end{acknowledgments}

\bibliography{biblio}

\begin{thebibliography}{36}
\expandafter\ifx\csname natexlab\endcsname\relax\def\natexlab#1{#1}\fi
\expandafter\ifx\csname bibnamefont\endcsname\relax
  \def\bibnamefont#1{#1}\fi
\expandafter\ifx\csname bibfnamefont\endcsname\relax
  \def\bibfnamefont#1{#1}\fi
\expandafter\ifx\csname citenamefont\endcsname\relax
  \def\citenamefont#1{#1}\fi
\expandafter\ifx\csname url\endcsname\relax
  \def\url#1{\texttt{#1}}\fi
\expandafter\ifx\csname urlprefix\endcsname\relax\def\urlprefix{URL }\fi
\providecommand{\bibinfo}[2]{#2}
\providecommand{\eprint}[2][]{\url{#2}}

\bibitem[{\citenamefont{Aad et~al.}(2012)}]{Aad:2012gk}
\bibinfo{author}{\bibfnamefont{G.}~\bibnamefont{Aad}} \bibnamefont{et~al.}
  (\bibinfo{collaboration}{ATLAS Collaboration}), \bibinfo{journal}{Phys.Lett}
  \textbf{\bibinfo{volume}{B716}}, \bibinfo{pages}{1} (\bibinfo{year}{2012}).

\bibitem[{\citenamefont{Chatrchyan et~al.}(2012)}]{Chatrchyan:2012gu}
\bibinfo{author}{\bibfnamefont{S.}~\bibnamefont{Chatrchyan}}
  \bibnamefont{et~al.} (\bibinfo{collaboration}{CMS Collaboration}),
  \bibinfo{journal}{Phys.Lett} \textbf{\bibinfo{volume}{B716}},
  \bibinfo{pages}{30} (\bibinfo{year}{2012}).

\bibitem[{\citenamefont{Nilles}(1984)}]{Nilles:1983ge}
\bibinfo{author}{\bibfnamefont{H.~P.} \bibnamefont{Nilles}},
  \bibinfo{journal}{Phys.Rept.} \textbf{\bibinfo{volume}{110}},
  \bibinfo{pages}{1} (\bibinfo{year}{1984}).

\bibitem[{\citenamefont{Haber and Kane}(1985)}]{Haber:1984rc}
\bibinfo{author}{\bibfnamefont{H.~E.} \bibnamefont{Haber}} \bibnamefont{and}
  \bibinfo{author}{\bibfnamefont{G.~L.} \bibnamefont{Kane}},
  \bibinfo{journal}{Phys.Rept.} \textbf{\bibinfo{volume}{117}},
  \bibinfo{pages}{75} (\bibinfo{year}{1985}).

\bibitem[{\citenamefont{{ATLAS
  Collaboration}}(2013{\natexlab{a}})}]{ATLAS-CONF-2013-049}
\bibinfo{author}{\bibnamefont{{ATLAS Collaboration}}}
  (\bibinfo{year}{2013}{\natexlab{a}}), \eprint{ATLAS-CONF-2013-049}.

\bibitem[{\citenamefont{{CMS
  Collaboration}}(2013{\natexlab{a}})}]{CMS-PAS-SUS-13-006}
\bibinfo{author}{\bibnamefont{{CMS Collaboration}}}
  (\bibinfo{year}{2013}{\natexlab{a}}), \eprint{CMS-PAS-SUS-13-006}.

\bibitem[{\citenamefont{Giudice and Rattazzi}(1999)}]{Giudice:1998bp}
\bibinfo{author}{\bibfnamefont{G.}~\bibnamefont{Giudice}} \bibnamefont{and}
  \bibinfo{author}{\bibfnamefont{R.}~\bibnamefont{Rattazzi}},
  \bibinfo{journal}{Phys.Rept.} \textbf{\bibinfo{volume}{322}},
  \bibinfo{pages}{419} (\bibinfo{year}{1999}).

\bibitem[{\citenamefont{Meade et~al.}(2009)\citenamefont{Meade, Seiberg, and
  Shih}}]{Meade:2008wd}
\bibinfo{author}{\bibfnamefont{P.}~\bibnamefont{Meade}},
  \bibinfo{author}{\bibfnamefont{N.}~\bibnamefont{Seiberg}}, \bibnamefont{and}
  \bibinfo{author}{\bibfnamefont{D.}~\bibnamefont{Shih}},
  \bibinfo{journal}{Prog.Theor.Phys.Suppl.} \textbf{\bibinfo{volume}{177}},
  \bibinfo{pages}{143} (\bibinfo{year}{2009}).

\bibitem[{\citenamefont{{CMS Collaboration}}(2013{\natexlab{b}})}]{SUS-13-002}
\bibinfo{author}{\bibnamefont{{CMS Collaboration}}}
  (\bibinfo{year}{2013}{\natexlab{b}}), \eprint{CMS-PAS-SUS-13-002}.

\bibitem[{\citenamefont{Ambrosanio et~al.}(1998)\citenamefont{Ambrosanio,
  Kribs, and Martin}}]{Ambrosanio:1997bq}
\bibinfo{author}{\bibfnamefont{S.}~\bibnamefont{Ambrosanio}},
  \bibinfo{author}{\bibfnamefont{G.~D.} \bibnamefont{Kribs}}, \bibnamefont{and}
  \bibinfo{author}{\bibfnamefont{S.~P.} \bibnamefont{Martin}},
  \bibinfo{journal}{Nucl.Phys.} \textbf{\bibinfo{volume}{B516}},
  \bibinfo{pages}{55} (\bibinfo{year}{1998}).

\bibitem[{\citenamefont{Ruderman and Shih}(2010)}]{Ruderman:2010kj}
\bibinfo{author}{\bibfnamefont{J.~T.} \bibnamefont{Ruderman}} \bibnamefont{and}
  \bibinfo{author}{\bibfnamefont{D.}~\bibnamefont{Shih}},
  \bibinfo{journal}{JHEP} \textbf{\bibinfo{volume}{1011}}, \bibinfo{pages}{046}
  (\bibinfo{year}{2010}).

\bibitem[{\citenamefont{Evans et~al.}(2007)\citenamefont{Evans, Morrissey, and
  Wells}}]{Evans:2006sj}
\bibinfo{author}{\bibfnamefont{J.~L.} \bibnamefont{Evans}},
  \bibinfo{author}{\bibfnamefont{D.~E.} \bibnamefont{Morrissey}},
  \bibnamefont{and} \bibinfo{author}{\bibfnamefont{J.~D.} \bibnamefont{Wells}},
  \bibinfo{journal}{Phys.Rev.} \textbf{\bibinfo{volume}{D75}},
  \bibinfo{pages}{055017} (\bibinfo{year}{2007}).

\bibitem[{\citenamefont{Grajek et~al.}(2013)\citenamefont{Grajek, Mariotti, and
  Redigolo}}]{Grajek:2013ola}
\bibinfo{author}{\bibfnamefont{P.}~\bibnamefont{Grajek}},
  \bibinfo{author}{\bibfnamefont{A.}~\bibnamefont{Mariotti}}, \bibnamefont{and}
  \bibinfo{author}{\bibfnamefont{D.}~\bibnamefont{Redigolo}},
  \bibinfo{journal}{JHEP} \textbf{\bibinfo{volume}{1307}}, \bibinfo{pages}{109}
  (\bibinfo{year}{2013}).

\bibitem[{\citenamefont{Beenakker et~al.}(1999)\citenamefont{Beenakker, Klasen,
  Kramer, Plehn, Spira et~al.}}]{Beenakker:1999xh}
\bibinfo{author}{\bibfnamefont{W.}~\bibnamefont{Beenakker}},
  \bibinfo{author}{\bibfnamefont{M.}~\bibnamefont{Klasen}},
  \bibinfo{author}{\bibfnamefont{M.}~\bibnamefont{Kramer}},
  \bibinfo{author}{\bibfnamefont{T.}~\bibnamefont{Plehn}},
  \bibinfo{author}{\bibfnamefont{M.}~\bibnamefont{Spira}},
  \bibnamefont{et~al.}, \bibinfo{journal}{Phys.Rev.Lett.}
  \textbf{\bibinfo{volume}{83}}, \bibinfo{pages}{3780} (\bibinfo{year}{1999}).

\bibitem[{\citenamefont{Bozzi et~al.}(2006)\citenamefont{Bozzi, Fuks, and
  Klasen}}]{Bozzi:2006fw}
\bibinfo{author}{\bibfnamefont{G.}~\bibnamefont{Bozzi}},
  \bibinfo{author}{\bibfnamefont{B.}~\bibnamefont{Fuks}}, \bibnamefont{and}
  \bibinfo{author}{\bibfnamefont{M.}~\bibnamefont{Klasen}},
  \bibinfo{journal}{Phys.Rev.} \textbf{\bibinfo{volume}{D74}},
  \bibinfo{pages}{015001} (\bibinfo{year}{2006}).

\bibitem[{\citenamefont{Bozzi et~al.}(2007)\citenamefont{Bozzi, Fuks, and
  Klasen}}]{Bozzi:2007qr}
\bibinfo{author}{\bibfnamefont{G.}~\bibnamefont{Bozzi}},
  \bibinfo{author}{\bibfnamefont{B.}~\bibnamefont{Fuks}}, \bibnamefont{and}
  \bibinfo{author}{\bibfnamefont{M.}~\bibnamefont{Klasen}},
  \bibinfo{journal}{Nucl.Phys.} \textbf{\bibinfo{volume}{B777}},
  \bibinfo{pages}{157} (\bibinfo{year}{2007}).

\bibitem[{\citenamefont{Bozzi et~al.}(2008)\citenamefont{Bozzi, Fuks, and
  Klasen}}]{Bozzi:2007tea}
\bibinfo{author}{\bibfnamefont{G.}~\bibnamefont{Bozzi}},
  \bibinfo{author}{\bibfnamefont{B.}~\bibnamefont{Fuks}}, \bibnamefont{and}
  \bibinfo{author}{\bibfnamefont{M.}~\bibnamefont{Klasen}},
  \bibinfo{journal}{Nucl.Phys.} \textbf{\bibinfo{volume}{B794}},
  \bibinfo{pages}{46} (\bibinfo{year}{2008}).

\bibitem[{\citenamefont{Fuks et~al.}(2013)\citenamefont{Fuks, Klasen, Lamprea,
  and Rothering}}]{Fuks:2013vua}
\bibinfo{author}{\bibfnamefont{B.}~\bibnamefont{Fuks}},
  \bibinfo{author}{\bibfnamefont{M.}~\bibnamefont{Klasen}},
  \bibinfo{author}{\bibfnamefont{D.~R.} \bibnamefont{Lamprea}},
  \bibnamefont{and}
  \bibinfo{author}{\bibfnamefont{M.}~\bibnamefont{Rothering}},
  \bibinfo{journal}{Eur.Phys.J.} \textbf{\bibinfo{volume}{C73}},
  \bibinfo{pages}{2480} (\bibinfo{year}{2013}).

\bibitem[{\citenamefont{Argurio et~al.}(2012)\citenamefont{Argurio,
  De~Causmaecker, Ferretti, Mariotti, Mawatari et~al.}}]{Argurio:2011gu}
\bibinfo{author}{\bibfnamefont{R.}~\bibnamefont{Argurio}},
  \bibinfo{author}{\bibfnamefont{K.}~\bibnamefont{De~Causmaecker}},
  \bibinfo{author}{\bibfnamefont{G.}~\bibnamefont{Ferretti}},
  \bibinfo{author}{\bibfnamefont{A.}~\bibnamefont{Mariotti}},
  \bibinfo{author}{\bibfnamefont{K.}~\bibnamefont{Mawatari}},
  \bibnamefont{et~al.}, \bibinfo{journal}{JHEP}
  \textbf{\bibinfo{volume}{1206}}, \bibinfo{pages}{096} (\bibinfo{year}{2012}).

\bibitem[{\citenamefont{Mawatari and Takaesu}(2011)}]{Mawatari:2011jy}
\bibinfo{author}{\bibfnamefont{K.}~\bibnamefont{Mawatari}} \bibnamefont{and}
  \bibinfo{author}{\bibfnamefont{Y.}~\bibnamefont{Takaesu}},
  \bibinfo{journal}{Eur.Phys.J.} \textbf{\bibinfo{volume}{C71}},
  \bibinfo{pages}{1640} (\bibinfo{year}{2011}).

\bibitem[{\citenamefont{Christensen and Duhr}(2009)}]{Christensen:2008py}
\bibinfo{author}{\bibfnamefont{N.~D.} \bibnamefont{Christensen}}
  \bibnamefont{and} \bibinfo{author}{\bibfnamefont{C.}~\bibnamefont{Duhr}},
  \bibinfo{journal}{Comput.Phys.Commun.} \textbf{\bibinfo{volume}{180}},
  \bibinfo{pages}{1614} (\bibinfo{year}{2009}).

\bibitem[{\citenamefont{Duhr and Fuks}(2011)}]{Duhr:2011se}
\bibinfo{author}{\bibfnamefont{C.}~\bibnamefont{Duhr}} \bibnamefont{and}
  \bibinfo{author}{\bibfnamefont{B.}~\bibnamefont{Fuks}},
  \bibinfo{journal}{Comput.Phys.Commun.} \textbf{\bibinfo{volume}{182}},
  \bibinfo{pages}{2404} (\bibinfo{year}{2011}).

\bibitem[{\citenamefont{Degrande et~al.}(2012)\citenamefont{Degrande, Duhr,
  Fuks, Grellscheid, Mattelaer et~al.}}]{Degrande:2011ua}
\bibinfo{author}{\bibfnamefont{C.}~\bibnamefont{Degrande}},
  \bibinfo{author}{\bibfnamefont{C.}~\bibnamefont{Duhr}},
  \bibinfo{author}{\bibfnamefont{B.}~\bibnamefont{Fuks}},
  \bibinfo{author}{\bibfnamefont{D.}~\bibnamefont{Grellscheid}},
  \bibinfo{author}{\bibfnamefont{O.}~\bibnamefont{Mattelaer}},
  \bibnamefont{et~al.}, \bibinfo{journal}{Comput.Phys.Commun.}
  \textbf{\bibinfo{volume}{183}}, \bibinfo{pages}{1201} (\bibinfo{year}{2012}).

\bibitem[{\citenamefont{Alwall et~al.}(2011)\citenamefont{Alwall, Herquet,
  Maltoni, Mattelaer, and Stelzer}}]{Alwall:2011uj}
\bibinfo{author}{\bibfnamefont{J.}~\bibnamefont{Alwall}},
  \bibinfo{author}{\bibfnamefont{M.}~\bibnamefont{Herquet}},
  \bibinfo{author}{\bibfnamefont{F.}~\bibnamefont{Maltoni}},
  \bibinfo{author}{\bibfnamefont{O.}~\bibnamefont{Mattelaer}},
  \bibnamefont{and} \bibinfo{author}{\bibfnamefont{T.}~\bibnamefont{Stelzer}},
  \bibinfo{journal}{JHEP} \textbf{\bibinfo{volume}{1106}}, \bibinfo{pages}{128}
  (\bibinfo{year}{2011}).

\bibitem[{\citenamefont{Sjostrand et~al.}(2006)\citenamefont{Sjostrand, Mrenna,
  and Skands}}]{Sjostrand:2006za}
\bibinfo{author}{\bibfnamefont{T.}~\bibnamefont{Sjostrand}},
  \bibinfo{author}{\bibfnamefont{S.}~\bibnamefont{Mrenna}}, \bibnamefont{and}
  \bibinfo{author}{\bibfnamefont{P.~Z.} \bibnamefont{Skands}},
  \bibinfo{journal}{JHEP} \textbf{\bibinfo{volume}{0605}}, \bibinfo{pages}{026}
  (\bibinfo{year}{2006}).

\bibitem[{\citenamefont{Jadach et~al.}(1993)\citenamefont{Jadach, Was, Decker,
  and Kuhn}}]{Jadach:1993hs}
\bibinfo{author}{\bibfnamefont{S.}~\bibnamefont{Jadach}},
  \bibinfo{author}{\bibfnamefont{Z.}~\bibnamefont{Was}},
  \bibinfo{author}{\bibfnamefont{R.}~\bibnamefont{Decker}}, \bibnamefont{and}
  \bibinfo{author}{\bibfnamefont{J.~H.} \bibnamefont{Kuhn}},
  \bibinfo{journal}{Comput.Phys.Commun.} \textbf{\bibinfo{volume}{76}},
  \bibinfo{pages}{361} (\bibinfo{year}{1993}).

\bibitem[{\citenamefont{Ovyn et~al.}(2009)\citenamefont{Ovyn, Rouby, and
  Lemaitre}}]{Ovyn:2009tx}
\bibinfo{author}{\bibfnamefont{S.}~\bibnamefont{Ovyn}},
  \bibinfo{author}{\bibfnamefont{X.}~\bibnamefont{Rouby}}, \bibnamefont{and}
  \bibinfo{author}{\bibfnamefont{V.}~\bibnamefont{Lemaitre}}
  (\bibinfo{year}{2009}), \eprint{0903.2225}.

\bibitem[{\citenamefont{Agram et~al.}(2013)\citenamefont{Agram, Andrea, Conte,
  Fuks, Gel\'e et~al.}}]{Agram:2013koa}
\bibinfo{author}{\bibfnamefont{J.-L.} \bibnamefont{Agram}},
  \bibinfo{author}{\bibfnamefont{J.}~\bibnamefont{Andrea}},
  \bibinfo{author}{\bibfnamefont{E.}~\bibnamefont{Conte}},
  \bibinfo{author}{\bibfnamefont{B.}~\bibnamefont{Fuks}},
  \bibinfo{author}{\bibfnamefont{D.}~\bibnamefont{Gel\'e}},
  \bibnamefont{et~al.}, \bibinfo{journal}{Phys. Lett.}
  \textbf{\bibinfo{volume}{B725}}, \bibinfo{pages}{123} (\bibinfo{year}{2013}).

\bibitem[{\citenamefont{Conte et~al.}(2013)\citenamefont{Conte, Fuks, and
  Serret}}]{Conte:2012fm}
\bibinfo{author}{\bibfnamefont{E.}~\bibnamefont{Conte}},
  \bibinfo{author}{\bibfnamefont{B.}~\bibnamefont{Fuks}}, \bibnamefont{and}
  \bibinfo{author}{\bibfnamefont{G.}~\bibnamefont{Serret}},
  \bibinfo{journal}{Comput.Phys.Commun.} \textbf{\bibinfo{volume}{184}},
  \bibinfo{pages}{222} (\bibinfo{year}{2013}).

\bibitem[{\citenamefont{Cacciari et~al.}(2008)\citenamefont{Cacciari, Salam,
  and Soyez}}]{Cacciari:2008gp}
\bibinfo{author}{\bibfnamefont{M.}~\bibnamefont{Cacciari}},
  \bibinfo{author}{\bibfnamefont{G.~P.} \bibnamefont{Salam}}, \bibnamefont{and}
  \bibinfo{author}{\bibfnamefont{G.}~\bibnamefont{Soyez}},
  \bibinfo{journal}{JHEP} \textbf{\bibinfo{volume}{0804}}, \bibinfo{pages}{063}
  (\bibinfo{year}{2008}).

\bibitem[{\citenamefont{Cacciari et~al.}(2012)\citenamefont{Cacciari, Salam,
  and Soyez}}]{Cacciari:2011ma}
\bibinfo{author}{\bibfnamefont{M.}~\bibnamefont{Cacciari}},
  \bibinfo{author}{\bibfnamefont{G.~P.} \bibnamefont{Salam}}, \bibnamefont{and}
  \bibinfo{author}{\bibfnamefont{G.}~\bibnamefont{Soyez}},
  \bibinfo{journal}{Eur.Phys.J.} \textbf{\bibinfo{volume}{C72}},
  \bibinfo{pages}{1896} (\bibinfo{year}{2012}).

\bibitem[{\citenamefont{Abbiendi et~al.}(2006)}]{Abbiendi:2005gc}
\bibinfo{author}{\bibfnamefont{G.}~\bibnamefont{Abbiendi}} \bibnamefont{et~al.}
  (\bibinfo{collaboration}{OPAL Collaboration}), \bibinfo{journal}{Eur.Phys.J.}
  \textbf{\bibinfo{volume}{C46}}, \bibinfo{pages}{307} (\bibinfo{year}{2006}).

\bibitem[{\citenamefont{{ATLAS
  Collaboration}}(2013{\natexlab{b}})}]{ATLAS-CONF-2013-028}
\bibinfo{author}{\bibnamefont{{ATLAS Collaboration}}}
  (\bibinfo{year}{2013}{\natexlab{b}}), \eprint{ATLAS-CONF-2013-028}.

\bibitem[{\citenamefont{D'Hondt et~al.}(to appear)\citenamefont{D'Hondt,
  De~Causmaecker, Fuks, Mariotti, Mawatari, Petersson, and
  Redigolo}}]{ToAppear}
\bibinfo{author}{\bibfnamefont{J.}~\bibnamefont{D'Hondt}},
  \bibinfo{author}{\bibfnamefont{K.}~\bibnamefont{De~Causmaecker}},
  \bibinfo{author}{\bibfnamefont{B.}~\bibnamefont{Fuks}},
  \bibinfo{author}{\bibfnamefont{A.}~\bibnamefont{Mariotti}},
  \bibinfo{author}{\bibfnamefont{K.}~\bibnamefont{Mawatari}},
  \bibinfo{author}{\bibfnamefont{C.}~\bibnamefont{Petersson}},
  \bibnamefont{and} \bibinfo{author}{\bibfnamefont{D.}~\bibnamefont{Redigolo}}
  (\bibinfo{year}{to appear}).

\bibitem[{\citenamefont{{CMS
  Collaboration}}(2013{\natexlab{c}})}]{CMS-PAS-SUS-13-010}
\bibinfo{author}{\bibnamefont{{CMS Collaboration}}}
  (\bibinfo{year}{2013}{\natexlab{c}}), \eprint{CMS-PAS-SUS-13-010}.

\bibitem[{\citenamefont{{ATLAS
  Collaboration}}(2013{\natexlab{c}})}]{ATLAS-CONF-2013-036}
\bibinfo{author}{\bibnamefont{{ATLAS Collaboration}}}
  (\bibinfo{year}{2013}{\natexlab{c}}), \eprint{ATLAS-CONF-2013-036}.

\end{thebibliography}

\end{document}